\documentclass[%
reprint,
 amsmath,amssymb,
 aps,
]{revtex4-1}
 \usepackage[colorlinks=true, allcolors=blue]{hyperref}
\usepackage{graphicx}
\usepackage{dcolumn}
\usepackage{bm}
\usepackage{lipsum}

\begin{document}

\title{Hole in the Deuteron}

\author{Misak Sargsian}

\affiliation{Florida International University, Miami, FL 33199}

\date{\today} 

\begin{abstract}
We introduce a new observable, $A_{node}$, that allows 
to isolate the node in the S-partial wave 
distribution of the deuteron in high $Q^2$ electro-disintegration processes 
with tensor polarized target. The node is a signature of nuclear repulsive core and 
represents a crucial test of  its strength, size as well as  the role 
of  the relativistic and non-nucleonic effects in  a deeply bound state. Within plane wave 
impulse approximation the node immitates a ``hole" through  which incoming 
probe passes  through without interaction. 
It is demonstrated that  high $Q^2$ electro-disintegration 
processes 
due to their strong anisotropy of 
final state interaction effects, allow to gain an unprecedented 
access to the node, opening up a new venue in probing elusive dynamics  of 
the nuclear repulsive core. 
\end{abstract}

\maketitle


\noindent{\bf Introduction:}
One of the most fascinating properties of nuclear forces is 
the nuclear repulsive core which provides a stability for  atomic nuclei, making 
it possible the emergence of a structure for the visible matter\cite{Wilczek:2007hco}.
Without it nuclei will collapse to  sizes $\sim 1$~fm,   followed by  the onset of quark-gluon degrees of freedom and restoration of chiral symmetry with rather unimaginable consequences for the order of the universe that we know.

\noindent{\bf History of the Problem:}
Already in 1950's it was observed\cite{Blatt:1952ije} 
that for the case of  attractive
two-nucleon forces the atomic nuclei with $A\sim 200$ will collapse to the distances half of the $NN$ interaction length, with per-nucleon binding energies 
$\sim 1600$~MeV\! (compare to actual 8MeV). Moreover no saturation density is possible, with the nuclear binding energy growing as $\sim A^2$.
Initially it was thought that the solution lies in the exchange character of nuclear forces as well as many-body effects\cite{Blatt:1952ije}.  
However the solution came  in 1950's, from the theoretical analysis of the  Berkeley 345~MeV pp scattering data which exhibited almost isotropic angular distribution  for $20^0- 90^0$ range of CM scattering angle. Such an isotropy was described by Jastrow\cite{Jastrow:1950zz,Jastrow:1951vyc}, by introducing a short range hard repulsive interaction, surrounded by an attractive well in the $^1S_0$ channel of pp interaction. The further analyses of pn  data demonstrated that similar  repulsion exists also in $^3S_1$ channel with the core distances estimated  to be $r_c\approx 0.4-0.6$~fm\cite{Arndt:1966in,Walecka:1995mi}.

\noindent{\bf Current Status:} After seven decades, the dynamical origin of the repulsive 
core is as elusive as ever, providing little  understanding  why atomic nuclei are stable.
The most modern $NN$ potentials based on the fitting of phase-shifts still use the phenomenological ansatz for the repulsive core introduced in 1960's.
Attempts to describe the core through vector-meson exchanges face conceptual issues, 
mainly, on how to describe $\le 0.6$~fm inter-nucleon distances by 
the exchange of mesons with comparable or larger radii\cite{Feynman:1973xc}.
The situation is not better in effective theories in which short distance 
dynamics of NN interaction are absorbed in contact terms which are evaluated 
by comparing the calculations with low energy observables.

With the emergence of quantum chromodynamics~(QCD), the studies of dynamical origin of the nuclear core obtained new dimensionality. QCD predicts robust inelastic transitions as well as 
sizable hidden color component in the NN system at the core distances\cite{Harvey:1981,Brodsky:1986}. One of the first attempts to explore 
the repulsive core in the S-channel of NN system was done within Lattice 
QCD\cite{Ishii:2006ec}, in which the obtained core confined at shorter distances ($\lesssim$ 0.5~Fm) and  is much softer compared to the phenomenological potentials. The latter may be the reflection of the fact that no inelastic and hidden color components were included in the calculation. Further theoretical progress in calculating the core will require an inclusion of the hidden component in the NN system as well as accounting for the inelastic transitions - both imitate a  repulsion due to orthogonality of these states with the 
detected $NN$ states  in the final state.

\noindent{\bf Probing the NN Core:}
The lack of progress is also related to the fact that there is very few processes 
(except perhaps the  ones in the cores of neutron stars) that are directly related  
the dynamics  of the nuclear core.
The advance in understanding of the dynamical origin of the nuclear core will require relevant experimental data to be confronted with  theoretical calculations. However at least three main challenges should be overcome in laboratory experiments  in order to probe the nuclear core; (i) Experiments should provide  sufficiently high energy and momentum transfer to reach the  sub-fermi ($\le 0.6$~fm) distances 
in the $NN$ system, (ii) Due to the nature of repulsion the experiments should be designed to measure diminishingly small cross sections, and (iii) Specific observables should be identified that are directly related to the dynamics of the nuclear core.

Never before the above three conditions were available in nuclear experiments. The situation is changed with 12~GeV energy upgrade of Jefferson Lab\cite{Dudek:2012vr}  which can 
provide high energy  and high intensity electron beam for nuclear experiments. 

The experimental approach in this case  is to probe the $^3S_1$ state in 
the deuteron or in short range $pn$ correlations in nuclei.  For  the $^1S_0$ 
state the repulsive core can be probed  in short-range $pp$ correlations in 
light nuclei. The methodology of these experiments is to consider high energy and momentum transfer electro-production reactions in which the detection of a struck nucleon from the deuteron (or struck and recoil nucleon from NN SRC) will allow 
to probe largest possible  internal momenta in the NN system. The expectations in these 
experiments are that for pn systems above $800$~MeV/c internal momenta, the 
$^3S_1$-state starts to dominate the D-state, therefore the measured momentum distribution 
will be sensitive to the strength  and the composition of the nuclear core. 
Indeed, the very first (and limited) measurement of high $Q^2$ break-up of the  deuteron at very high missing momenta, $p_m$,\cite{HallC:2020kdm} observed that starting at $p_m\ge 750$~MeV/c, the obtained momentum distribution is in qualitative  
disagreement with the predictions based on conventional  wave functions of the deuteron.
This can be the first indication of  possible non-nucleonic components in the ground state wave function of the deuteron\cite{Sargsian:2022rmq} that starts to dominate   above the inelastic threshold of the isosinglet pn system.
The discussed reaction is currently the most promising process in reaching 
the nuclear core and the analysis of the new data\cite{Boeglin:2014aca}
may provide important new results.  However the full theoretical analysis 
in this case, will require a 
reliable description of the $D-$ wave at large internal momenta due to its 
contribution being large or comparable to the $S-$ partial wave in the deuteron.

The alternative approach presented in this work, is to isolate S-wave contribution in the deuteron through the measurement of new kind of observable, referred to as $A_{node}$ by
using polarized deuteron target.  The  $A_{node}$ should exhibit  a node
in the momentum distribution directly  related to the dynamics of the repulsive core and as calculations show can be measured  in exclusive deuteron break-up
reactions at large $Q^2$.

\begin{figure}[hbt]
\vspace{-0.34cm}
    \centering
    \includegraphics[width=6cm,height=2.9cm]{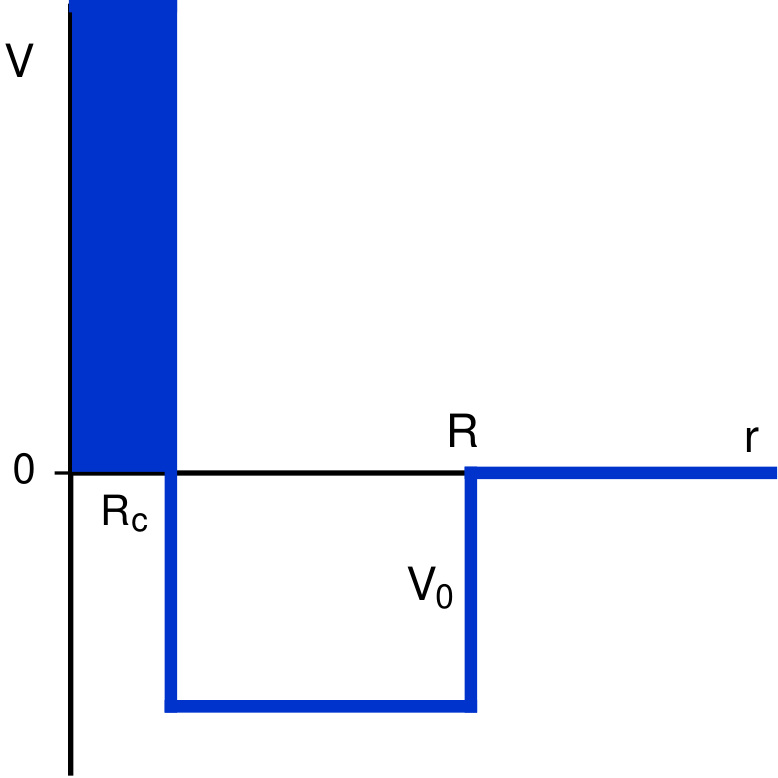}
        \caption{Simple spherical well potential, with radius $R$ and strength $V_0$,
        with infinite repulsive core at $r\le R_0$. }
    \label{fig:potential}
\end{figure}

\noindent {\bf The NN Core and the Node in the S-State Distribution in the Deuteron:} 
We start with the simplistic  discussion of the momentum distribution for $^3S_1$ bound state in the deuteron in which potential is described  (similar to Refs.\cite{Jastrow:1950zz,Walecka:1995mi}) as a spherical well of radius $R$ with infinite repulsive core at the center with radius $R_c$(Fig.\ref{fig:potential}). 

\begin{figure}[hbt]
\vspace{-0.4cm}
   \centering 
    \includegraphics[width=9.0cm,height=4cm]{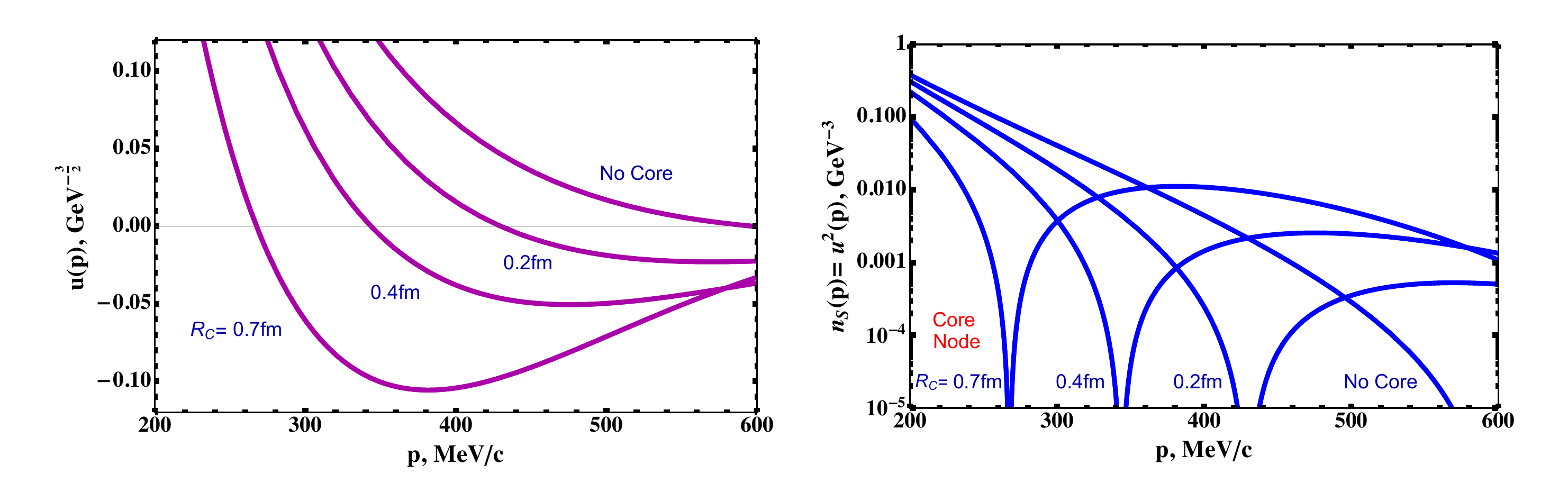}
        \caption{(left) $^3S_1$- partial wave (left)  and  its momentum distribution (right)  
        calculated  for different sizes of 
        the $pn$ core.}
    \label{fig:u_and_n}
\end{figure}

In this case we consider not the pp scattering as in Ref.\cite{Jastrow:1950zz} but bound pn $^3S_1$-state with  energy of the state matched to the  deuteron bound state.
Using Schroedinger equation we calculate the radial wave function, $R_S(r)$  and then  construct its momentum distribution as:
\begin{equation}
n(p) = |u(p)|^2,  \ \mbox{where} \  \ u(p) = {2\over \sqrt{2\pi}}\int j_0(pr) R_s(r)rdr.
\label{smomdist}
\end{equation}
For the potential of Fig.\ref{fig:potential}, the  infinite core solution is 
$R_S(r) = A \sin{(k_0(R-R_C)})$ for $R_c\le r \le R$ and 
$R_S(e) = Be^{-\kappa r}$, for $r> R$, where $k_0 = \sqrt{M(|E_b|+V_0)}$ and $\kappa = \sqrt{M|E_b|}$, with $M$ and $E_b$ being 
reduced mass and deuteron binding energy respectively. Applying continuity condition for the wave function and its first derivative and then using Fourier transform of Eq.(\ref{smomdist}) one obtains the momentum space radial wave function presented in Fig.\ref{fig:u_and_n} (left panel) at different radii of the core. 

As it is seen in the figure the momentum space wave function changes its sign,  which  is due to the $\sin$ function in the coordinate space with the  boundary condition of the hard core $R_S(R_c)=0$.  The crossing zeros in $R_S(r)$ correspond to the roots of transcendental equation with approximate relation for the crossing momentum:
\begin{equation}
p_0 \approx  k_0  ( 1 - {R_C\over R})  + {\pi\over R}
(1 - {R_C\over R})^2.
\end{equation}
The crossing  zero for the  wave function in the momentum space will 
result in a node in the momentum distribution defined according to Eq.\ref{smomdist} (Fig.\ref{fig:u_and_n}  (right panel)). These nodes are the reflection of the nuclear core 
that in the considered model are related the core's size.
\medskip

Momentum distribution is  a quantity which  enters  in the 
cross section of exclusive $d(e,e^\prime N_f)N_r$ processes in the 
plane wave impulse approximation (PWIA).  In PWIA  electron 
knocks out  one of the nucleons from the deuteron without further re-interactions in the final state.  
If  deuteron consisted of only  the $S$-state,  
then in this case the node  is like a {\em hole} in the momentum space  through which 
the probe-electron will pass without interaction. 

\begin{figure}[hbt]
   \centering
    \includegraphics[width=8.cm,height=4cm]{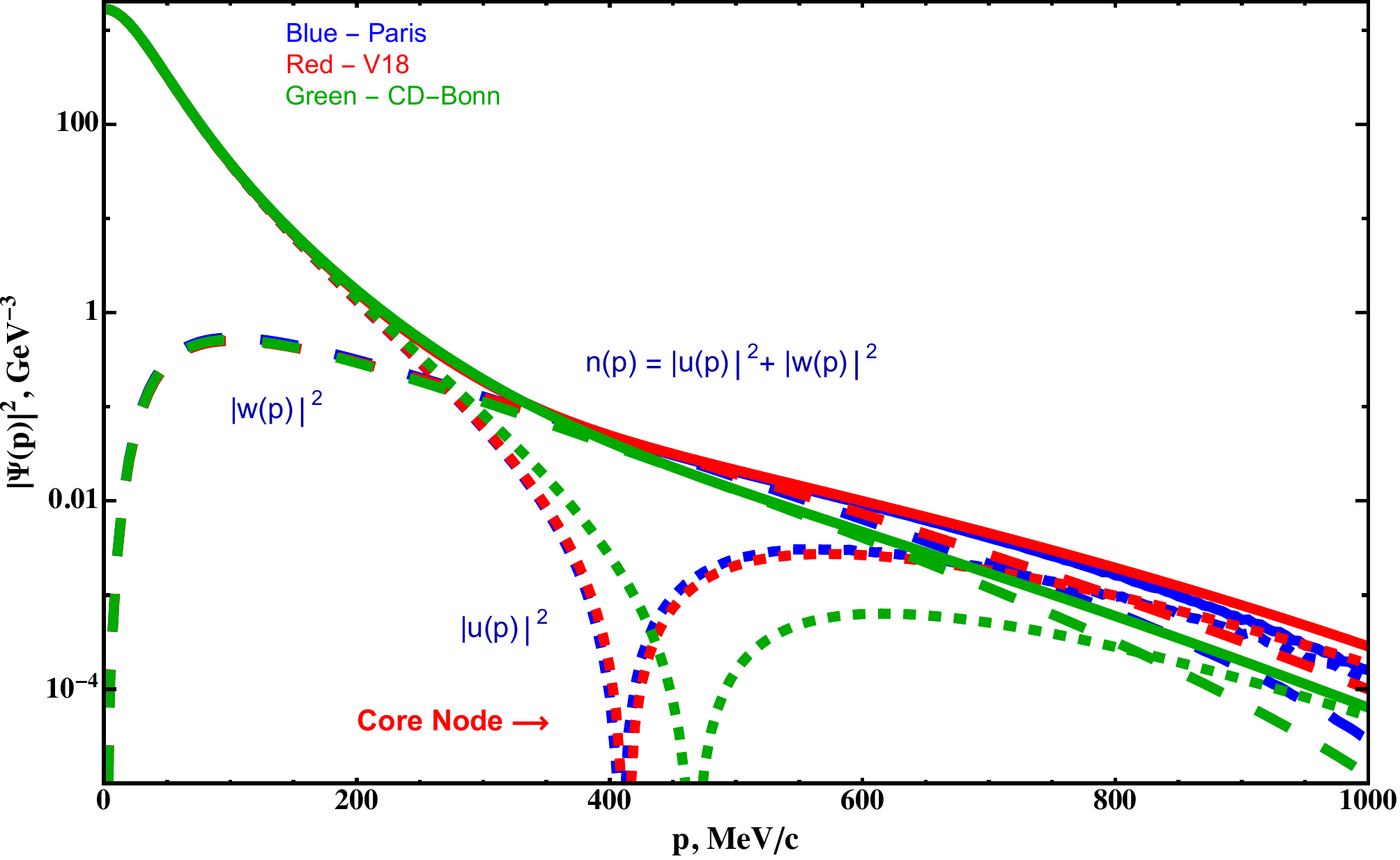}
        \caption{ Momentum distribution of partial $S$ (dotted) and $D$ (dashed)
        waves in the deuteron, as well as total momentum distribution (solid) for 
        wave functions calculated with Paris, V18 and CD-Bonn potentials.}
    \label{fig:deuteron}
\end{figure}

To see whether discussed simplistic model has a relevance for 
the realistic deuteron, in Fig.\ref{fig:deuteron} we present 
the momentum distribution of the nucleon in the deuteron calculated with realistic Paris\cite{Lacombe:1981eg}, V18\cite{Wiringa:1994wb} (hard) and CD-Bonn\cite{Machleidt:2000ge} (soft) NN potentials. 

As the figure shows the S-wave momentum distribution indeed has a node and its position is related to the ``hardness" of  the NN potential. It is known that CD-Bonn potential is more soft predicting lesser high momentum component than the one with Paris and V18 potentials.
However the realistic deuteron wave function also has a $D$- partial wave, $w(p)$ and the actual  momentum distribution is a sum of both S- and D- wave distributions: $n(p)= |u(p)|^2 + |w(p)|^2$. Because of this, as Fig.\ref{fig:deuteron} shows, Nature choose to hide the node of the  S-wave momentum distribution by the robust D-wave distribution in the  region of interest.

{\bf Isolating S-wave Contribution:} The deuteron momentum distribution theoretically can be measured  in above mentioned $d(e,e^\prime, N_f)N_r$ reactions from unpolarized 
deuteron target within PWIA,  in which  it  probes 
unpolarized deuteron density matrix:
\begin{equation} 
    \rho_{unp}(p)= n(p) = |u(p)|^2 + |w(p)|^2.
    \label{rhounp}
\end{equation}
As Fig.\ref{fig:deuteron} shows in this case to  be sensitive to the  NN core one needs to measure very large internal momenta $\ge 800$~MeV/C in which case S-wave distribution becomes comparable or exceeds that D-wave distribution. However, 
in this case, probing the core in the $^3S$-channel will not be direct due to the permanence of the $D$-wave contribution.

In this work an alternative approach  is suggested 
which allows to maximally isolate $S$-state contribution by 
assuming simultaneous measurement of $d(e,e^\prime, N_f)N_r$
reaction from unpolarized and tensor polarized deuteron.

The idea of using tensor polarized deuteron to probe the 
high momentum characteristics of the deuteron is more than half a century old. For the case of  $d(e,e^\prime, N_f)N_r$ reaction with tensor polarized deuteron it will probe 
the density matrix of the form:
\begin{eqnarray}
& & \rho_{20}(p,\theta_N)   \equiv  {|\psi_d^{1}|^2 + |\psi_d^{-1}|^2 - 2 |\psi_d^{0}|^2\over 3}    =  \ \ \ \  \nonumber \\ & & \ \ \ \ \ \ \   {3\cos^2(\theta_N)-1\over 2} 
\left[ 2\sqrt{2} u(p)w(p) - w^2(p)\right],
\label{rho20}
\end{eqnarray}
in which the $S$- partial wave contribution is suppressed 
(no $|u(p)|^2$ term) allowing to explore the properties of $D$-partial wave. 
Here  $\psi_d^m$- represents the deuteron wave function with polarization $m=-1,0,1$ and $\theta_N$ is the direction of internal momenta with respect to 
the polarization axis of the deuteron.

Combining now Eqs.(\ref{rhounp}) and (\ref{rho20}) in the form:
\begin{eqnarray}
\rho_{node}(p) & = &   \rho_{unp}(p)  + {2\rho_{20}\over 3\cos^2(\theta_N)- 1} =   \nonumber \\
& = & u^2(p) + 2\sqrt{2} u(p)w(p), 
\label{rhonode}
\end{eqnarray}
one achieves an opposite effect in which $|w(p)|^2$ does not enter and the combination will have a node corresponding to the crossing zero of the $S$-state wave function in momentum space.

It is more practical to consider a new  asymmetry, $A_{node}$ defined as:
\begin{eqnarray}
A_{node}(p) \equiv {\rho_{node}(p) \over \rho_{unp}(p)} & = &  
1 + {2A_{zz}(p,\theta_N) \over 3\cos^2(\theta_N)- 1} \nonumber \\ 
& = &   {u^2(p) + 2\sqrt{2} u(p)w(p)\over u(p)^2 + w(p)^2},
\label{Anode}
\end{eqnarray}
where $A_{zz} = {\rho_{20}\over \rho_{unp}}$, which within PWIA
corresponds to the ratio of cross sections of $d(e,e^\prime, N_f)N_r$ scattering from tensor polarized and unpolarized deuteron target:
\begin{equation}
    T_{20} = {\sigma_{tensor}\over \sigma_{unp}}.
\label{T20}
\end{equation}

\begin{figure}[ht]
    \centering
    \includegraphics[width = 0.7\linewidth]{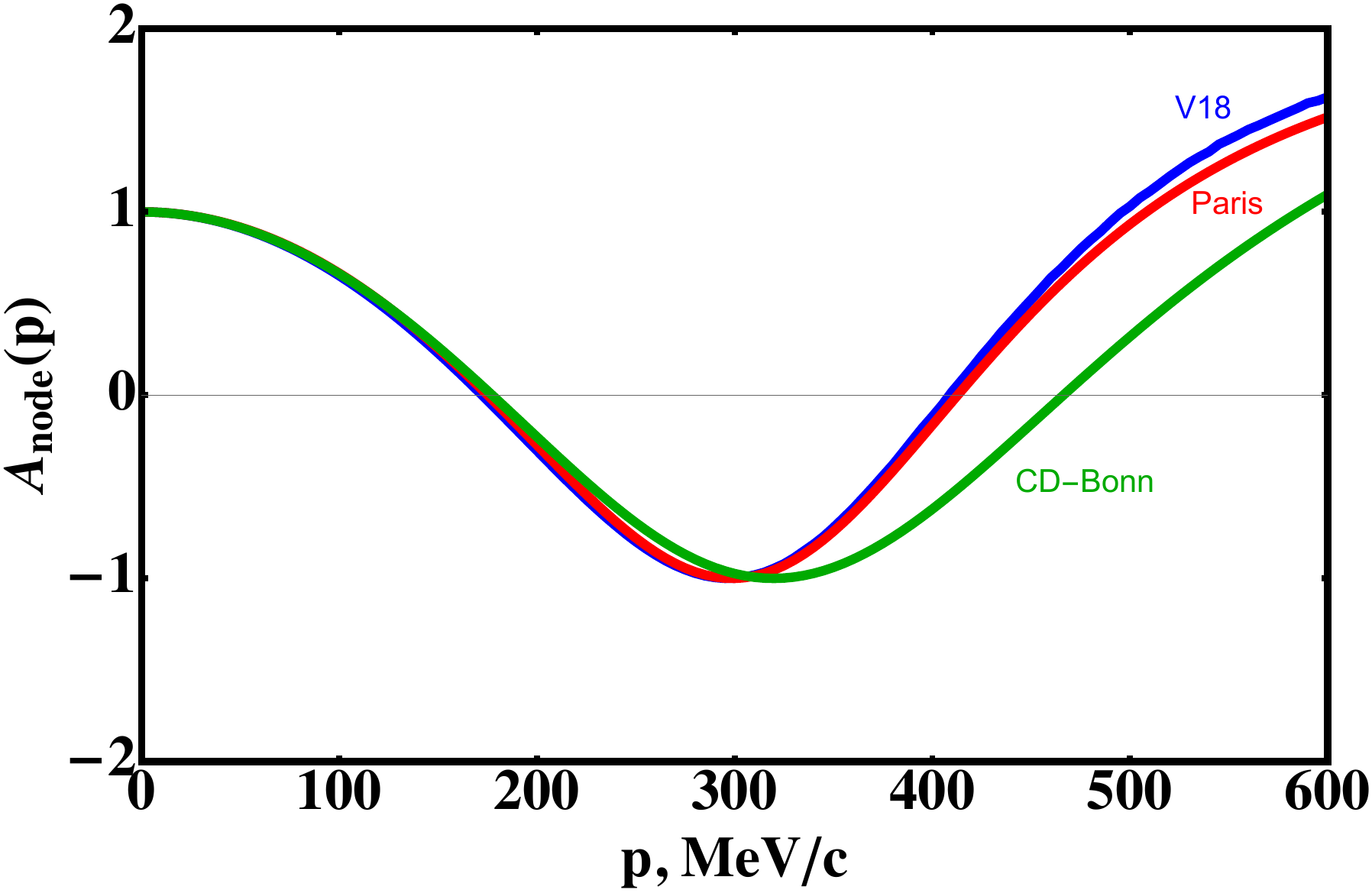}
\vspace{-0.3cm}
        \caption{Calculations   $A_{node}(p)$ using deuteron wave function calculated with Paris (red), CD-Bonn (green) and V18 (blue) potentials.}
    \label{fig:asym-dist1}
\end{figure}

In Fig.\ref{fig:asym-dist1} asymmetry $A_{node}(p)$ is calculated for $\theta_{N} = 0$. 
It  shows large sensitivity of  the  node positions at  $p\ge 400$~MeV/c,
to the choice of the NN potentials with which deuteron wave  functions 
are calculated. This is consistent with S-state momentum distributions in Fig.\ref{fig:deuteron}.

Note that $A_{node}$ has another zero 
at $p\approx 180$~MeV/c which corresponds to the condition of 
\begin{equation}
u(p) = - 2\sqrt{2} w(p).
\end{equation}
Since the above relation takes place at relatively small 
internal moments for which  the deuteron wave function is well known, the observation of this node will help to 
calibrate the experimental measurement of $A_{node}$.

\noindent
{\bf Feasibility of the  $A_{node}(p)$ Measurement:} Few decades ago the above discussion on possibility of isolating the S-state of the deuteron would have been purely theoretical, since low energy $d(e,e^\prime, N_f)N_r$ experiments with $Q^2< 1$~GeV$^2$ were completely dominated by non-PWIA contributions such as recoil production of final nucleon, Final State Interactions (FSI), Meson-Exchange currents~(MEC) and Isobar Contribution(IC)\cite{Arenhovel:1982rx,Boeglin:2015cha}. As a result it was impossible to establish direct connection between internal momentum of the deuteron and measured momenta of 
scattered electron and  nucleon, $N_f$.

\begin{figure}[ht]
    \centering
    \includegraphics[width = 0.8\linewidth]{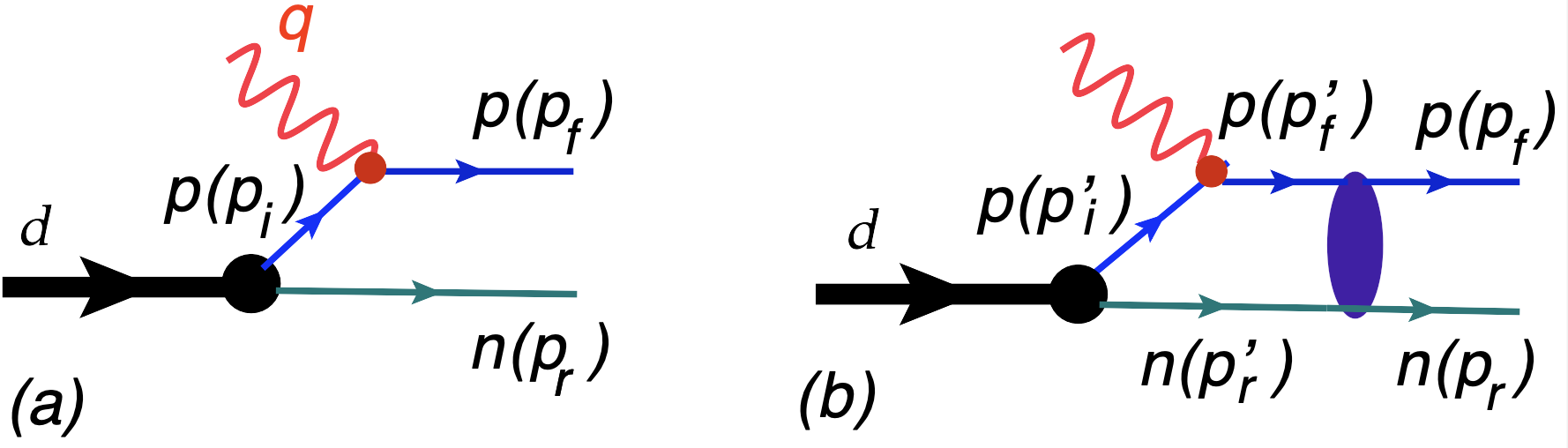}\\ 
\vspace{0.2cm}
    \includegraphics[width=6.4cm,height=4cm]{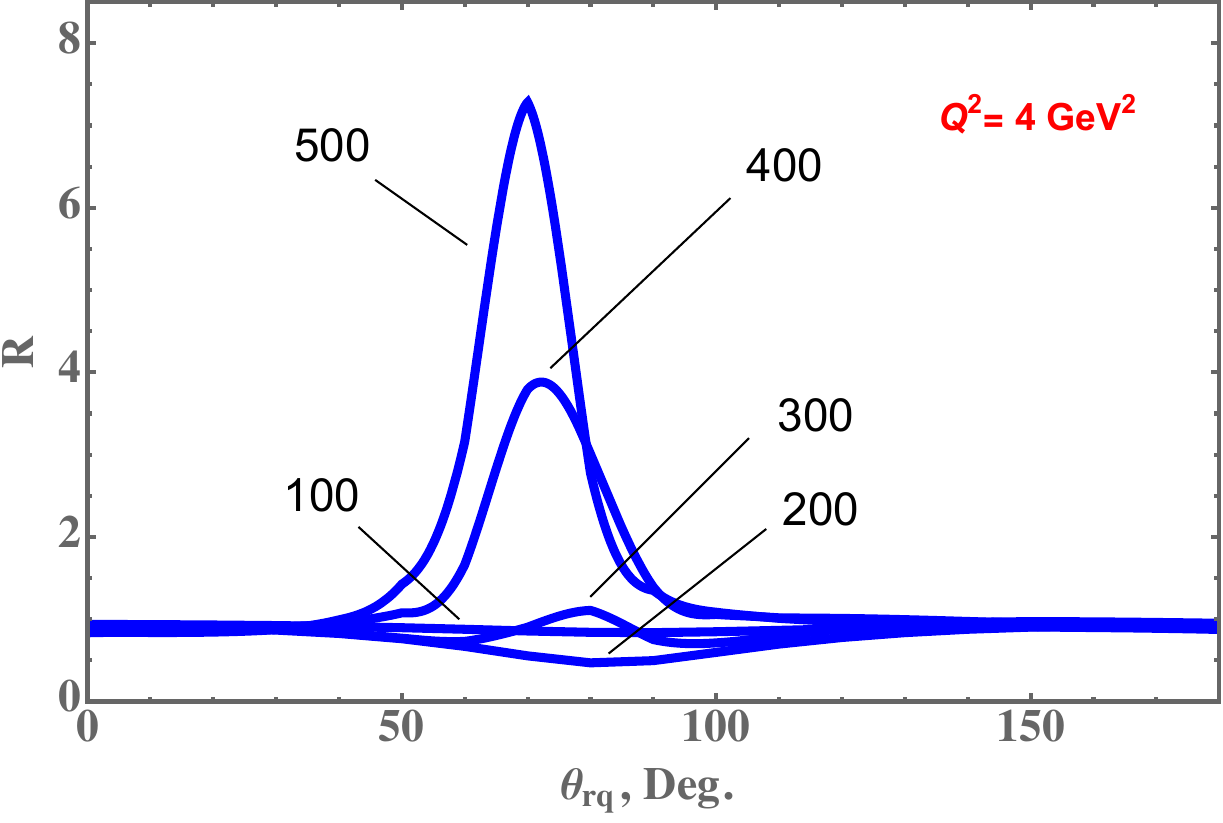}
    \vspace{-0.3cm}
        \caption{(Upper panel) (a) PWIA and (b) direct FSI contributions
to $d(e,e^\prime, N_f)N_r$ reaction in eikonal approximation. (Lower panel) The 
$\theta_{rq}$ dependence of the ratio (R)  of calculated cross section with FSI to  the  PWIA cross section,  for different values of missing momenta at $Q^2 = 4$~GeV$^2$.
    }
    \label{fig:gea}
\end{figure}

This situation changed significantly with the  emergence of high energy and intensity electron beam at Jefferson Lab which made it possible to measure 
 at $Q^2> few $~GeV$^2$.  In this case, choosing momentum of the final nucleon comparable with the transfered large momentum, ${\bf q}$ it was possible to identify clearly the final nucleon as the one which was struck by the virtual photon. The large $Q^2$ allows to suppress MEC contribution and charge-interchangee FSI, and finally the possibility of measuring at the kinematics of Bjorken $x>1$ (away from inelastic threshold) allowed to suppress the intermediate state Isobar contribution.
As a result in high $Q^2$ only PWIA  and direct FSI processes contribute to the $d(e,e^\prime, N_f)N_r$ reaction,
Fig.\ref{fig:gea} (upper panel). 

Another important advantage of high energy kinematics is the 
onset of the eikonal regime of the scattering. There was an  
intensive theoretical research in calculating the $d(e,e^\prime, N_f)N_r$ reaction in eikonal approximation\cite{Frankfurt:1996xx,Sargsian:2001ax,Jeschonnek:2008zg,Laget:2004sm,CiofidegliAtti:2004jg,Sargsian:2009hf,Ford:2013uza}. 
One of such approaches was the generalized eikonal approximation~(GEA) which developed  an
effective Feynman diagrammatic rules for  calculation of the nuclear scattering amplitude.  Within GEA one calculates all contributions\cite{Frankfurt:1996xx,Sargsian:2001ax,Sargsian:2009hf} to the nuclear scattering 
amplitude. For the simplicity of discussion we present the GEA predictions for  dominant amplitudes only, which are PWIA, $A_0^\mu$ and single 
rescattering, $A_1^\mu$ terms:
\vspace{-0.2cm}
\begin{eqnarray}
& & \hspace{-1.4cm}
\langle s_f,s_r \mid A_{0}^\mu\mid s_d\rangle  =  \sqrt{2}\sqrt{(2\pi)^3 2 E_r}\times \nonumber \\
& & \ \ \ \ \sum\limits_{s_i} J_{N}^\mu(s_f,p_f;s_i,p_i)
\Psi_d^{s_d}(s_i,p_i,s_r,p_r), 
\label{A0}
\end{eqnarray}
\vspace{-0.6cm}
\begin{eqnarray}
& & \hspace{-0.8cm} \langle s_f,s_r \mid A_{1}^\mu\mid s_d\rangle   = 
\nonumber \\
& & \hspace{-0.5cm}  =   
{i \sqrt{2}(2\pi)^{3\over 2}\over 4} \sum\limits_{s^\prime_f,s^\prime_r,s_i} \int {d^2p_r^\prime \over  (2\pi)^2}  
\frac{\sqrt{s(s-4m^2)}} {\sqrt{2\tilde E^\prime_r} |q|}    \nonumber \\
& & \times\langle p_f,s_f;p_r,s_r\mid f^{NN,on}(t,s)\mid \tilde p^\prime_r,s^\prime_r; \tilde p^\prime_f,s^\prime_f\rangle \nonumber \\
& &   \times J_{N}^\mu(s^\prime_f,p^\prime_f;s_i,\tilde p^\prime_i) \cdot  \Psi_d^{s_d}(s_i,\tilde p^\prime_i,s^\prime_r,\tilde p^\prime_r) 
\nonumber  \\
& & -  {(2\pi)^{3\over 2}\over \sqrt{2}} \sum\limits_{s^\prime_f,s^\prime_r,s_i} {\cal P}\int {dp^\prime_{r,z}\over 2\pi} 
\int {d^2p_r^\prime \over  (2\pi)^2} 
\frac{\sqrt{s(s-4m^2)}} {\sqrt{2E^\prime_r} |{\bf q}|} 
\nonumber \\
& & \times{\langle p_f,s_f;p_r,s_r\mid f^{NN,off}(t,s)\mid p^\prime_r,s^\prime_r;p^\prime_f,s^\prime_f\rangle
\over  p^\prime_{r,z}- \tilde p^\prime_{r,z} } \nonumber \\
& & \times J_{N}^\mu(s^\prime_f,p^\prime_f;s_i,p^\prime_i)
\cdot  \Psi_d^{s_d}(s_i,p^\prime_i,s^\prime_r,p^\prime_r),
\label{A1}
\end{eqnarray}
where all momenta are defined in Fig.\ref{fig:gea}. Here, $s = (p_f+ p_r)^2$, 
$m$ is the nucleon mass, $\tilde p^\prime_r = (p_{r,z}-\Delta, p^\prime_{r,\perp})$, $\tilde E^\prime_r = \sqrt{m^2 + \tilde p^{\prime 2}_r}$, 
$\tilde p^\prime_i = p_d - \tilde p^\prime_r$ and $\tilde p^\prime_f = \tilde p^\prime_i + q$ and 
\begin{equation}
\Delta = {q_0\over |{\bf q}|}(E_r - E^\prime_r) + {M_d\over |{\bf q}|}(E_r - E^\prime_r)  + {p^{\prime 2}_r-m^2\over 2  |{\bf q}|}.
\label{Delta}
\end{equation}
Above, $J_N^\mu$ is the electromagnetic current 
of bound nucleon calculated within virtual nucleon approximation\cite{Sargsian:2009hf} and 
the deuteron wave function is defined 
at initial relative momenta of $pn$ system. The on-shell part of the 
$pn$ scattering amplitude, $f^{NN,on}(t,s)$ is taken from experiments, while
the off-shell counterpart is modeled as follows:
\begin{equation}
f^{NN,off} = f^{NN,on} e^{B(m_{off}^2 -m^2)},
\label{fnn_off}
\end{equation}
where $m_{off}^2 \equiv (p^\prime_f)^2$.
Note that the rendered uncertainty due to  Eq.(\ref{fnn_off}) is very small 
since principal value part of the rescattering amplitude is parameterically small.

With $A_0^\mu$ and $A_1^\mu$ the differential cross section 
for  $s_d$ polarized deuteron can be written as follows
\begin{eqnarray}
&&\hspace{-0.4cm}
\frac {d\sigma^{s_d}}{dE_e^\prime, d\Omega_{e^\prime} d p_f d\Omega_f} = 
\frac { \alpha^2E^\prime_e}{Q^4 E_e} \times
\nonumber \\
&&
\hspace{-0.4cm}
 {1\over 2} \sum\limits_{s_f,s_r,s_1,s_2} \frac{\mid J_e^\mu J_{d,\mu}\mid^2}{2M_d E_f}{p_f^2\over 
\mid {p_f\over E_f} + {p_f - q cos(\theta_{p_f,q})\over E_r}\mid}
\label{crs}
\end{eqnarray}
where $E_e$ and $E^\prime_e$ are energies of the incoming and scattered electrons, $\alpha$ is the Fine Structure constant. Here $J_e^\mu$ is the leptonic current and $J_d^\mu$ is the electromagnetic transition current of the polarized  deuteron defined as
\begin{equation}
J_d^\mu = { \langle s_f,s_r \mid A_0^\mu + A_1^\mu\mid s_d\rangle\over \sqrt{2(2\pi)^3 2 E_r}}.
\label{Jd}
\end{equation}

The presented GEA approach successfully described the  data 
emerging from high $Q^2$ experiments with unpolarized deuteron targets from JLAB,  
for recoil nucleon momenta up to 550~MeV/c (see e.g. Refs.\cite{Egiyan:2007qj,HallA:2011gjn,HallC:2020kdm,Boeglin:2024spd}).  
One of the important results in theoretical analysis of 
these data is  the confirmation of strong angular anisotropy of FSI effects in 
which  the rescatteirng is concentrated at the transverse kinematics of recoil nucleon 
angle with respect to the direction of momentum transfer $q$, $60^0 \le \theta_{rq} \le 90^0$ (Fig.\ref{fig:gea}).   As Fig.\ref{fig:gea} (lower panel) shows, where $R$ is the ratio of calculated cross section with $A_0^\mu + A_1^\mu$ to the 
PWIA cross section,  
this anisotropy   can be used to identify kinematics most optimal for 
probing internal structure of the deuteron with small correction due to FSI effects. 
This corresponds to the $\theta_{min} < \theta_{rq} < 50^0$, where $\theta_{min}$ is defined 
by the minimal value of the relative momentum in the final $pn$ system for which eikonal 
approximation is valid. The suppression of FSI in this region is  confirmed by comparison with 
the large $Q^2$ data\cite{HallA:2011gjn,HallC:2020kdm,Boeglin:2024spd}. It is worth mentioning that similar  criteria can be used also for 
semi-exclusive processes involving other targets\cite{Sargsian:2005ru}.

\begin{figure}[t]
    \centering
    \includegraphics[width = 0.7\linewidth]{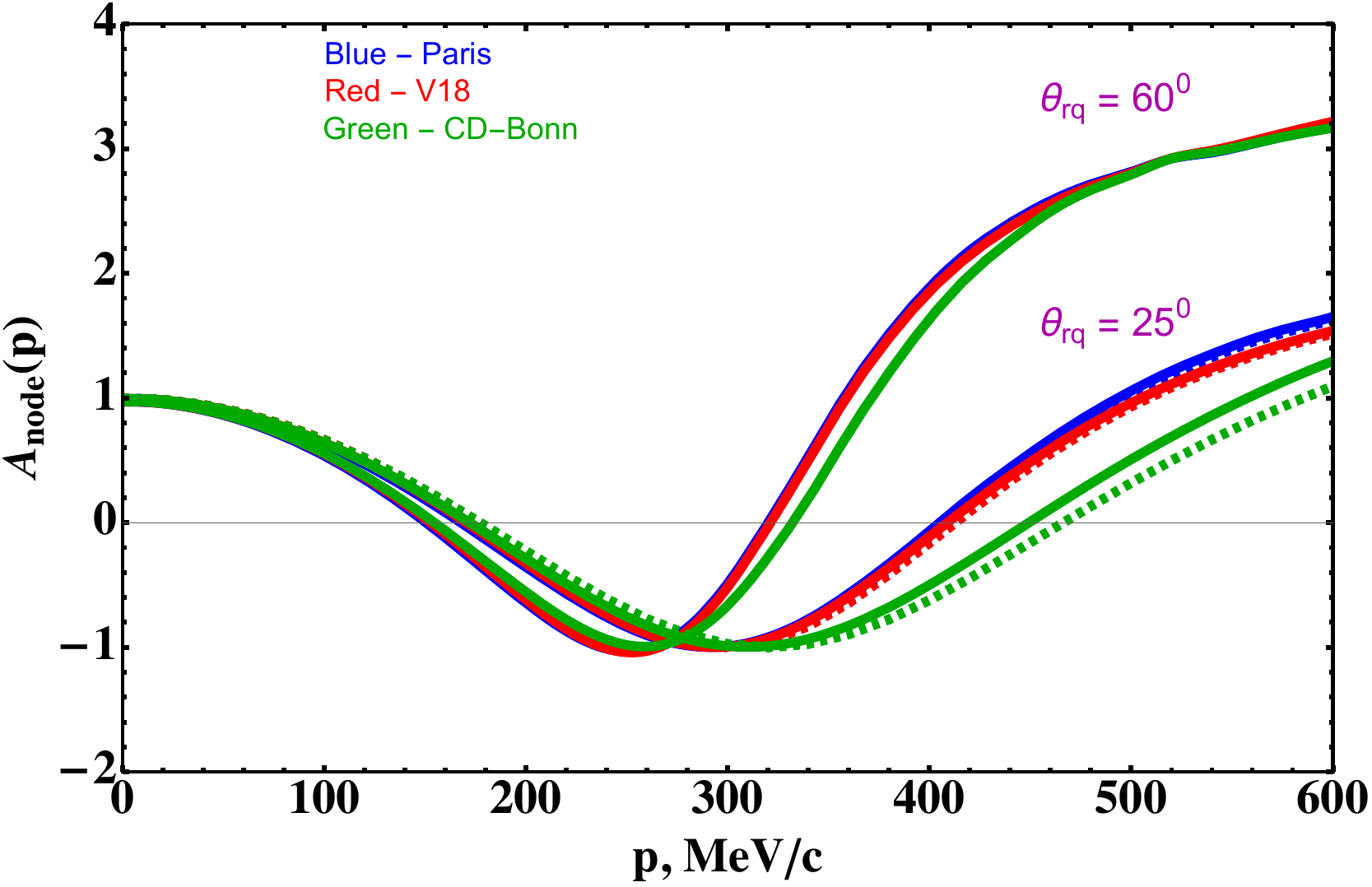}
\vspace{-0.4cm}
        \caption{Calculations   $A_{node}(p)$ using deuteron wave functions calculated with Paris (blue), CD-Bonn (green) and V18 (red) potentials for $Q^2=4$~GeV$^2$ and $\theta_{rq} = 
        60^0$ and $\theta_{rq}=25^0$ recoil angles. Solid line for calculation including FSI, dashed line PWIA calculation.}
    \label{fig:asym_GEA}
\end{figure}

In Fig.\ref{fig:asym_GEA} the calculation of $A_{node}$ is presented for $Q^2=4$~GeV$^2$ for 
kinematics of large and small FSI.
As the figure shows for the kinematics of large FSI ($\theta_{rq}= 60^0$) different wave functions predict close values for  $A_{node}$, which is expected due to the fact that FSI 
amplitude is dominated by small initial momenta for which different deuteron wave
functions are similar.  However for suppressed FSI kinematics ($\theta_{rq}= 25^0$) the results are close to the PWIA prediction thus allowing direct  probe of the node related to the repulsive core in the $^3S_1$ channel.

\noindent
{\bf Summary and Outlook:} We suggested new observable $A_{node}$ that isolates $^3S_1$ state in 
the deuteron whose node in the momentum distribution is related to the strength of $pn$-repulsive core.  The study of such a structure becomes  possible with the emerging reality of high momentum transfer deuteron electro-disintegration  reactions for which eikonal regime of FSI allows 
to isolate kinematics dominated by PWIA.  Such studies will use also the development of new technologies which allow to employ polarized deuteron target with intensive electron beams\cite{Slifer:2013vma,Accardi:2023chb}.

It is worth mentioning that our goal was not to discuss which model of deuteron wave function or (including relativistic and non-nucleonic effects) are valid but to demonstrate that it is possible to isolate the $^3S_1$-$pn$ state that will allow direct exploration of the repulsive core. Such an experiment is 
currently feasible albeit expansive, however no alternative  experiments exist or planned for addressing decades long problem of understanding the nature of the nuclear repulsive core.

\noindent {\bf Acknowledgment:}
This work is supported by United States DOE
grant under contract  DE-FG02-01ER41172.

\bibliography{TheCoreReferences.bib}

\end{document}